# *Unleashing HHG Efficiency: The Role of Driving Pulse Duration*

Robert Klas[1,2,3*], Martin Gebhardt[1,2], Jan Rothhardt[1,2,3] and Jens Limpert[1,2,3]

[1]*Institute of Applied Physics, Abbe Center of Photonics, Friedrich-Schiller-Universität Jena, Albert-Einstein-Str. 15, 07745 Jena, Germany*

[2]*Helmholtz-Institute Jena, Fröbelstieg 3, 07743 Jena, Germany*

[3]*Fraunhofer Institute for Applied Optics and Precision Engineering, Albert-Einstein-Str. 7, 07745 Jena, Germany*

*Corresponding author: robert.klas@uni-jena.de

## Abstract

High harmonic generation (HHG) is a crucial technology for compact, high-brightness extreme ultraviolet (XUV) and soft X-ray sources, which are key to advancing both fundamental and applied sciences. The availability of advanced driving lasers, with tunable wavelength, power, and pulse duration, opens new opportunities for optimizing HHG-based sources. While scaling laws for wavelength are well understood, this work focuses on how pulse duration impacts HHG efficiency and introduces a unified framework that links microscopic dynamics to macroscopic performance. We establish a practical scaling law for the single-atom dipole moment under phase-matching conditions, demonstrating a $\tau^{-1}$ dependence at 515 nm wavelength. By connecting this microscopic scaling to macroscopic conversion efficiency, we provide clear guidelines for optimizing HHG output across different gases and driving wavelengths. Furthermore, we identify fundamental constraints, including the carrier-envelope-phase (CEP) walk-off, which limits efficiency at longer driver wavelengths and becomes especially significant for very short pulses. All predictions are based on simple, accessible formulas, eliminating the need for complex numerical simulations. Experiments confirm these predictions and highlight when short pulses are advantageous, particularly in scenarios where CEP walk-off and absorption effects are minimized. These findings offer practical principles for designing next-generation HHG sources, capable of Watt-level average power and extended spectral reach, enabling more versatile and powerful HHG-based XUV and soft X-ray sources.

## 1 Introduction

Ultrashort-pulsed and spatially coherent extreme ultraviolet (XUV) und soft X-ray light sources (1 nm to 100 nm) are a powerful tool to expand the frontiers of knowledge in fields as diverse as physics, chemistry, biology, and material sciences. They already enable to gain a deep element-selective understanding of matter on the atomic length- and timescale [1–7]. In this field of research, a key parameter for future experiments is a high brightness, for example to minimize measurement times, enhance signal to noise ratios or enable the scanning over large parameter spaces while using light with a high degree of spatial coherence.

Among the numerous spatially coherent XUV and soft X-ray sources developed to date, different complementary approaches have been established [8]: large scale facilities, such as free-electron lasers and synchrotrons [9,10], and table-top alternatives, such as laser-driven higher-order harmonic generation (HHG) [11] or soft X-ray lasers [12]. While large-scale facilities deliver extraordinary brightness values, their limited access and high cost in building and maintaining them are drawbacks. In contrast, table-top sources are compact, portable, and highly accessible. HHG-based systems, in particular, are widely used in laboratories around the world due to the relatively low complexity of the required laser systems and the absence of debris. These systems already deliver brightness levels high enough to drive a vast number of experiments that were previously only possible at large-scale facilities. Further increases in brightness will be beneficial for photon-hungry applications in science and, ultimately, in industrial applications. Due to the nature of the HHG process[11], the generated radiation is nearly diffraction-limited. Therefore, an increase in brightness can mostly be achieved by increasing the photon flux.

A proven approach is to increase the average power of the driving laser. This can be done by directly using a high average power laser in a single pass geometry [13], or by recycling the driving laser in an enhancement cavity [14]. In both cases, highest XUV-average powers in the milliwatt regime have been demonstrated [15,16]. This strategy is supported by the rapid advancements in high-average power laser technology. Here, ultrashort pulsed systems with average powers of up to and beyond 1 kW have been already shown [17–19]. This is an order of magnitude higher average power than used for state of the art HHG-beamlines [7,13,20], giving the possibility to increase the photon flux by one order of magnitude.

The second approach is to optimize the conversion efficiency of the HHG-process in a single pass. This can be done by utilizing short wavelength driving lasers or shorter driving pulse durations. The scaling of the efficiency with the fundamental wavelength can be considered as well-known [21]. However, although it has been experimentally shown, that ultrashort driving pulse durations lead to higher HHG efficiencies [13,22,23], qualitative and wholistic scaling laws of the efficiency with the driving pulse duration have not yet been derived and proven.

This paper investigates the scaling of HHG efficiency with the driving pulse duration, considering both single-emitter dynamics and the role of phase matching in systems with multiple coherent emitters. The derived scaling laws and limitations for ultrashort driving pulses provide general guidelines for selecting optimal laser parameters, offering a foundation for designing highly efficient HHG beamlines.

In the first part, a simple and quantitative formula for the single-emitter HHG efficiency is derived and experimentally validated. The analysis reveals that shorter pulse durations allow for the use of higher peak intensities[13], leading to an inverse scaling of the single-emitter HHG efficiency with the pulse duration. This formula provides an intuitive framework for predicting a conversion efficiency based on laser pulse parameters.

The second part examines the effects of carrier-envelope phase (CEP) walk-off caused by group velocity mismatches in the nonlinear medium. This phenomenon, where differences between the phase and group velocities of the driving laser lead to dephasing of the generated XUV radiation, has been studied theoretically, showing its prominence at longer driving wavelengths and shorter pulse durations [24,25]. Building on this foundation, the relevance of these effects is explored experimentally and theoretically over a broad range of pulse durations and wavelengths critical for high-efficiency HHG. It is shown that dephasing imposes a limit on the effective medium length, which is compared to the absorption length. This comparison allows for a detailed evaluation of the impact of CEP walk-off and provides insights into optimizing HHG efficiency under various experimental conditions.

## 2 Scaling of the high harmonic efficiency with the driving pulse duration

The theoretical concept for the scaling of the high harmonic generation efficiency with respect to the driving pulse duration is illustrated in Figure 1. This concept can be understood in two sequential steps: (1) the microscopic scaling of the single-atom dipole moment of an individual emitter, and (2) the macroscopic scaling of the coherent emission from a multitude of emitters. These two mechanisms will be discussed in detail in the following.

## Microscopic scaling in theory

At first, the scaling of the single atom dipole moment with the intensity of the driving field is discussed. From this it is possible to draw conclusions for the scaling of the HHG-efficiency with the driving pulse duration (red box in Figure 1).

In the first approximation, the single atom dipole moment ($|d_q|^2$) scales nonlinearly with the incident intensity $I_0$. The scaling is still discussed controversially in the community and ranges from $|d_q|^2 \propto I_0^{1 \text{ to } 10}$ in the plateau region [15,26–31]. Furthermore, the scaling depends on the fundamental wavelength and the nonlinear medium[31]. However, most application relevant publications that include supporting experiments [15,26,30,31], use the scaling that can be extracted by exponential fitting of Fig. 3 in the paper of Lewenstein et al. [27]:

$$|d_q|^2 \propto \left(\frac{I_0}{I_{q,\min}}\right)^{4.6}, \quad I_0 \geq I_{q,\min}, \qquad \qquad 1$$

while $I_{q,\min}$ is the minimum intensity needed to generate a photon with an energy equal to the q$^{\text{th}}$ harmonic. If the intensity is too low ($I_0 < I_{q,\min}$), i.e. in beyond the cutoff, the single-atom dipole moment is virtually zero[30]. Experimentally, equation 1 reveals that the most efficient generation of plateau harmonics is at the highest available intensity, i.e., at the temporal peak of the driving laser pulse.

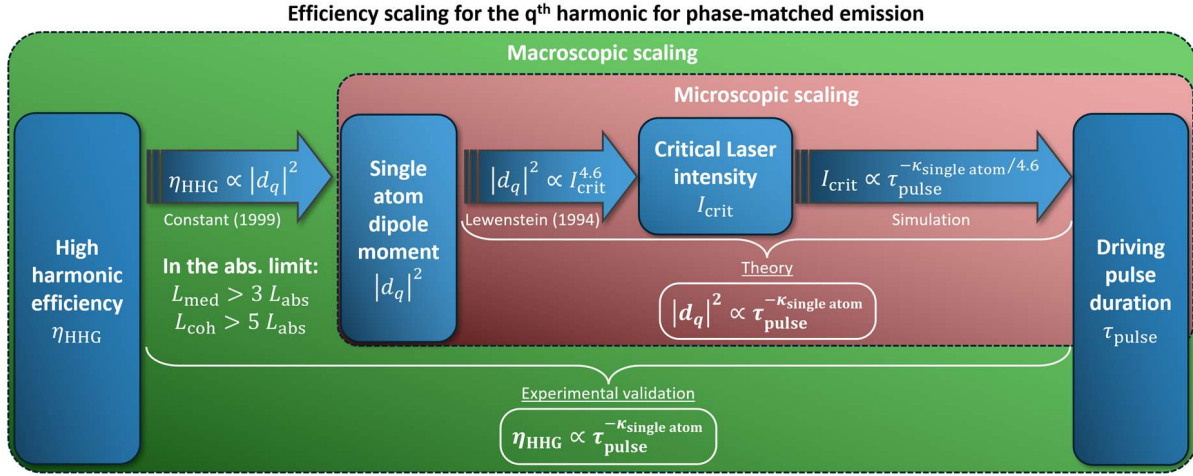

*Figure 1: Principle of the theoretical considerations for the pulse duration dependent scaling of the microscopic and macroscopic high harmonic efficiency for phase-matched emission.*

However, phase matching is crucial for efficient HHG[32]. The time-dependent contributions of dispersion from neutral atoms and plasma are particularly important[11], as true phase matching cannot be achieved beyond a certain ionization level, known as the critical ionization, depending on the wavelength of the driving laser and the gas type [32]. To achieve phase matching at the temporal peak of the laser pulse, i.e., at the highest available intensity, the intensity must be increased to a value where critical ionization is reached close to the temporal peak of the laser pulse. This specific intensity is referred to as the critical intensity [33]. Plugging the critical intensity into the cutoff-equation[34], results in the so-called phase matching cutoff[32]. The critical intensity increases with decreasing pulse durations (shown in Figure 2 a)). This is because equal ionization levels at the peak of the pulse can be reached using shorter pulse durations while applying significantly higher peak intensities[13,35]. Consequently, this results in the well-known fact of a higher phase matching cutoff when using shorter driving pulse durations[13].

The dependence of the critical intensity on the fundamental laser wavelength and pulse duration is non-trivial. However, the critical intensity can be calculated numerically using ADK[36] or PPT[37] ionization rates and iteratively check at what intensity the ionization at the temporal peak of the pulse is equal to the critical ionization [33]. An exemplary calculation using a fundamental wavelength of 515 nm is shown in Figure 2 a). Here, the critical intensity, depicted as solid lines, is calculated for different noble gases versus the pulse durations.

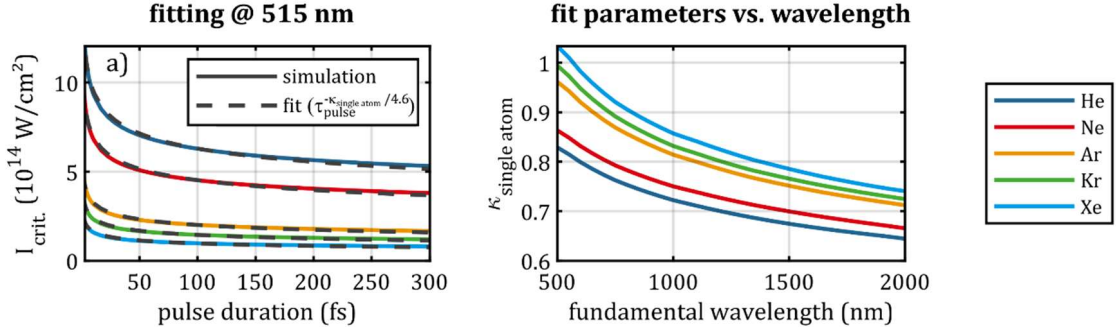

Figure 2: a) Pulse duration dependent critical intensity at central wavelength of 515 nm for different noble gases. The dashed lines correspond to power fits with the fitting parameter $\kappa_{single\ atom}$ for the different noble gases (He: 0.84, Ne: 0.87, Ar: 0.98, Kr: 1.02, Xe: 1.05). b) Wavelength dependence of the fitting parameter $\kappa_{single\ atom}$ for different noble gases.

At varying pulse durations $\tau_{pulse}$ and a fixed fundamental wavelength, the critical intensity can be approximated with a power-function. This is exemplarily shown as the well-coinciding dashed lines in Figure 2 a). Finally, this leads to a simple scaling law for the critical intensity $I_{crit}$:

$$I_{crit} \propto \tau_{pulse}^{-\kappa_{single\ atom}/4.6}, \quad\quad 2$$

with the scaling parameter $\kappa_{single\ atom}$ for the scaling of single atom dipole moment at a fixed wavelength and gas type with the driving pulse duration (Figure 2 b)).

Finally, a scaling law of the single atom dipole moment with the pulse duration for plateau harmonics can be derived by plugging equation 2 into equation 1 (with $I_0 = I_{crit}$):

$$|d_q|^2 \propto \tau_{pulse}^{-\kappa_{single\ atom}}. \quad\quad 3$$

The scaling parameter $\kappa_{single\ atom}$ primarily depends on the fundamental wavelength and the gas used, as shown in Figure 2(b). Most application relevant HHG-based sources use driving lasers with a fundamental wavelength of 500 nm to 2 µm[7]. For this spectral range $\kappa_{single\ atom}$ decreases with increasing wavelength. Additionally, at a given wavelength, it decreases with increasing ionization potentials. Ultimately, $\kappa_{single\ atom}$ is a positive number ranging from 0.6 to 1.1, indicating the beneficial increase in HHG efficiency with decreasing pulse duration, which is more pronounced at shorter fundamental wavelengths. For instance, using ten times shorter pulses at 500 nm results in a tenfold increase in efficiency, compared to a fivefold increase at 2 µm. To conclude this analysis, this finding confirms the statement that shorter pulses lead to higher macroscopic HHG efficiencies [38], adding quantitative information.

## Macroscopic scaling: experimental evidence

In the following, an experimental proof is provided, showing that the approximations that have been applied during the derivation of equation 3 are well-justified.

To experimentally validate equation 3, six independent experiments are compared [15,39–44]. All these experiments use a 515 nm driving laser for HHG in argon and reported an optimized efficiency (i.e. phase matching and optimal intensity) for a photon energy of 21.7 eV (9th harmonic) and 26.5 eV (11th harmonic). Furthermore, it has been carefully verified that the harmonics are in the plateau region (Table 1, supplement). The main difference in the experiments is the pulse energy (50 µJ to 450 µJ), the repetition rate (4 kHz to 334 kHz) and the average power of the driving laser systems (1.8 W to 30 W). However, at this parameter range, no detrimental effects such as accumulated ionization [16] or thermal effects on the optics [45] are expected or reported. Henceforth, the HHG efficiency can be considered independent on the pulse energy and repetition rate [46,47].

Furthermore, all presented experiments were conducted in the absorption-limited regime[26], as carefully verified in table S1 (supplement). This regime allows to gain deeper understanding of the single-atom dipole moment. In the absorption limit, the macroscopic HHG efficiency depends solely on the single-atom dipole moment $|d_q|^2$ and the absorption cross-section $\sigma$ of the used gas, and scales as $\eta_{\text{HHG}} \propto \left|\frac{d_q}{\sigma}\right|^2$ [26]. Using equation 3, this results in a scaling of the macroscopic and absorption limited HHG efficiency with the driving pulse duration of:

$$\eta_{\text{HHG}} \propto \tau_{\text{pulse}}^{-\kappa_{\text{single atom}}}. \qquad 4$$

In conclusion, these experiments are an excellent choice to investigate the efficiency scaling laws of the single atom yield with the driving pulse duration. For better statistics, the efficiency into two harmonic lines (9th and 11th) are evaluated and depicted in Figure 3 a). Overall, shorter driving pulse durations yield a higher HHG efficiency, as expected qualitatively[38].

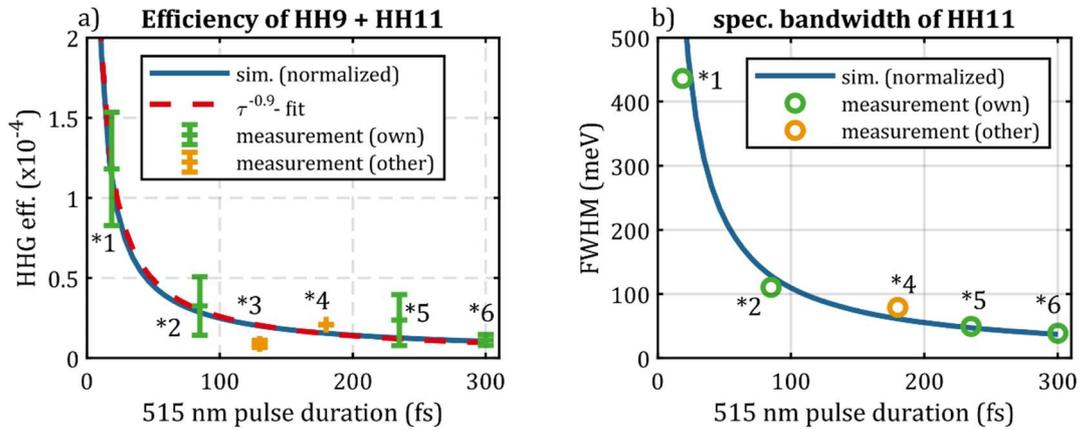

*Figure 3: HHG efficiency (a) for the generation of the 9th and 11th harmonic and spectral width (b) of the 11th harmonic of a 515 nm driven HHG source using argon at different driving pulse durations. The blue curves correspond to a phase matching simulation. The red dotted line is a power fit to the simulated data. Green and orange points are experimentally obtained results corresponding to measurements done by the authors and by other groups, respectively (\*1 [15,44], \*2 [39], \*3 [42], \*4 [43], \*5 [40], \*6 [41]). Figure adapted from [48].*

The theoretical scaling parameter for 515 nm driven HHG in argon is $\kappa_{\text{single atom}} = 0.98$ (Figure 2). Henceforth, the macroscopic and absorption limited efficiency of the 9th and 11th harmonic is theoretically

$$\eta_{q=9,11}(515 \text{ nm, argon}) \propto \tau^{-0.98}. \qquad 5$$

The estimation of HHG efficiency is validated by an additional simulation that accounts for phase matching effects and the varying number of laser cycles contributing to harmonic emission at different pulse durations. This simulation is based on the work of Constant et al.[26] and is detailed in the supplement of [15]. In comparison to the experiments, the medium length (430 µm), focal spot size (36 µm), and distance to the nozzle orifice (144 µm) are kept constant in the simulation. At each pulse duration, the pulse energy is adjusted to ensure that the intensity corresponds to the critical intensity. This adjustment is justified, as HHG efficiency is independent of pulse energy[47]. The resulting optimal phase matching pressure is consistently around 400 mbar, leading to an absorption length of 49 µm (56 µm) at 21.7 eV (26.5 eV), ensuring that all simulation conditions are within the absorption limit. This simulation does not calculate absolute efficiencies; therefore, the results have been normalized to align with the measured data.

The resulting dependency of the simulation is depicted as the blue curve in Figure 3 a) and is fitted with an exponential fit (red dashed line), showing a scaling of the efficiency with $\tau^{-0.9}$. Furthermore, the simulated phase matching window (i.e., the number of cycles contributing to the harmonic emission) can be used to calculate the bandwidth of the harmonic line via Fourier transformation. Figure 3 b) shows that the simulated bandwidths of the 11$^{th}$ harmonic fit well with the experimental data, confirming the validity of the simulation.

In conclusion, the two independent theoretical simulations fit well with the experimentally measured efficiencies, showing that the proposed scaling law is true for 515 nm in argon.

## 3 Macroscopic Effects: CEP walk-off-Limited High Harmonic Generation

The transfer of the microscopic scaling of the single atom dipole moment to a macroscopic HHG efficiency can be spoiled due to propagation effects that can lead to a decrease of the effective medium length and ultimately to non-absorption limited generation conditions. This is showcased in the following experiment, where HHG was driven by a laser with a central wavelength of 1030 nm, using pulses of either 30 fs or 7 fs. Detailed experimental procedures are provided in [35]. The resulting measured HHG efficiencies in the plateau region are presented in Figure 4.

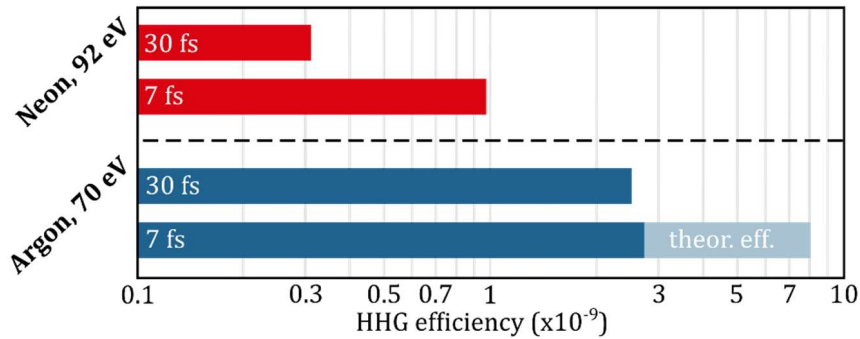

*Figure 4: Experimentally measured HHG efficiency generated with a 1030 nm laser at different pulse durations in neon (red) and argon (blue). The light blue bar is the theoretically expected single emitter efficiency (eq. 3).*

First the results obtained with neon are discussed (Figure 4 a)). The experiment is using a 150 µm medium length with a phase matching particle density that results in an absorption length of 32 µm at 92 eV[35]. Hence, the experiment is conducted in the absorption limited regime, which allows to gain information of the single atom efficiency scaling (equation 4). For this, the efficiency at a photon energy of 92 eV is evaluated, resulting in a 3.1 times higher efficiency for 7 fs versus 30 fs driven HHG.

This corresponds to a $\tau^{-0.78}$ scaling. The theoretical estimation of the single atom dipole moment and hence the macroscopic and absorption limited efficiency for a neon and central wavelength of 1030 nm yields a scaling of $\tau^{-0.75}$ (Figure 2 b)), showing an excellent agreement between theory and experiment.

In contrast to the measurements in neon, the efficiencies of plateau harmonics in argon at 70 eV are similar at different pulse durations. This contradicts the expectations of the absorption limited HHG-efficiency scaling of $\eta_{\text{HHG}} \propto \tau^{-0.8}$ (Figure 2 b)), highlighted as the light blue bar in Figure 4. The experimentally optimized medium length is 700 µm, while the used particle density results in an absorption length of 1 mm ($L_{\text{med}} = 0.7\, L_{\text{abs}}$). Even though the Rayleigh length is >4 mm, a further increase in medium length did not show any increase of the XUV photon flux. Hence, it is fundamentally impossible to reach the absorption limit.

The reason for the reduced effective medium length is caused by significant differences in the phase and group velocities of the driving laser pulse, leading to carrier-envelope-phase (CEP) walk-off. Theoretical investigations have shown that this effect is significant at ultrashort pulse durations and 1 µm or longer driving wavelengths [24,25,49]. Until now, the implications of a limited buildup length and a comparison to the absorption-limited regime have not been addressed. The goal of the next section is to study the macroscopic HHG efficiency for different driving pulse durations. Ultimately, this analysis will determine whether CEP walk-off is detrimental for specific experimental setups and will aid in the development of novel, highly efficient XUV and soft X-ray sources.

## Vivid Concept

First, the concept of CEP walk-off limitation is vividly described using the acceleration step of the classical equivalent for the high harmonic process, depicted in Figure 5. In the following simulation, a 7 fs pulse at a central wavelength of 1030 nm and an intensity of $4 \cdot 10^{14}\ \text{W/cm}^2$ is used without loss of generality. Additionally, only short trajectories with a kinetic energy of 90 eV at the time of return are considered.

Due to the dispersion of the plasma and the Gouy phase, the phase velocity of the driving laser exceeds its group velocity. This scenario is depicted in Figure 5, where the electrical field of driving laser pulse propagating through the ionized medium is shown at several positions (represented by different colors). For this illustration the framework of the phase velocity is used, due to its importance for phase matching. During propagation, the highlighted cycle (red line in b), c), and d)) from 0 to 3.4 fs, which should be phase matched, changes its position relative to the temporal peak and deforms due to the slower group velocity compared to the phase velocity.

Exemplarily, the short trajectories for electrons with a return energy of 90 eV are calculated for the red-highlighted half-cycle of the pulse at different positions. The colors of the trajectories correspond to the color of the respective electric field of the laser. The return time of 90 eV electrons in the laser field at the beginning (orange) and end (blue) of the medium length differ by up to 450 as. This is about ten times the cycle time of a 90 eV photon (45 as). Due to the uniformly continuous change of the electric field during propagation, there are 90 eV electrons with every return time in between, resulting in an incoherent superposition of the different XUV fields. To achieve a coherent superposition, only very small differences in the return time are permissible (e.g., 45 as in this specific example).

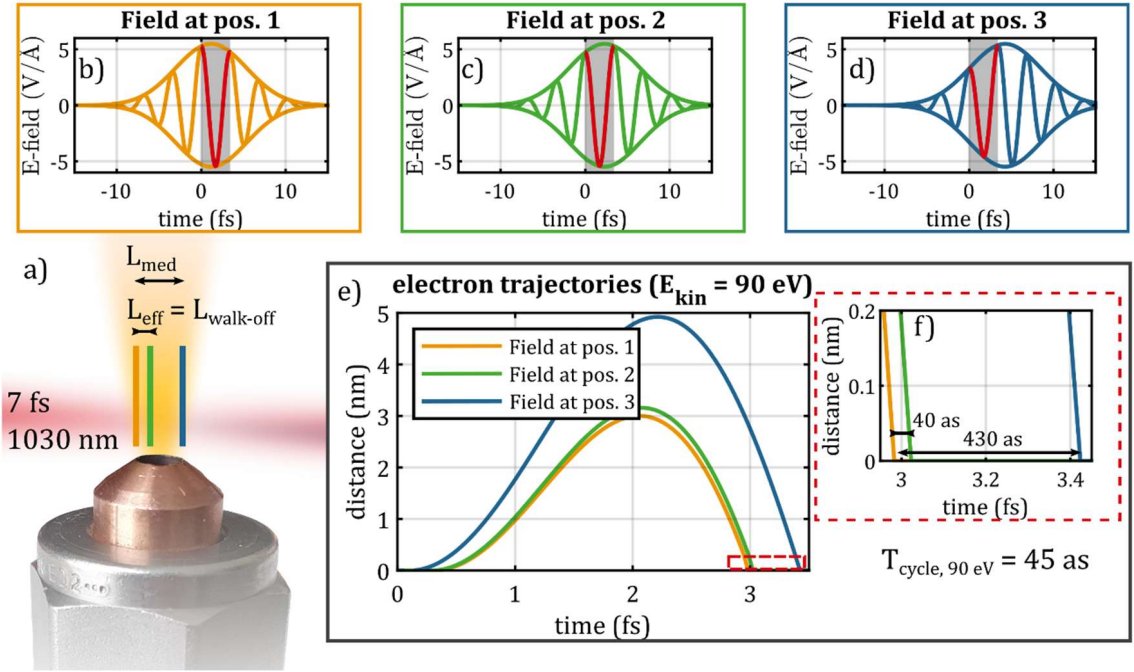

*Figure 5: Schematic principle of the CEP- walk-off limit. Without loss of generality the propagation of a 1030 nm, 7 fs pulse with an intensity of $4 \cdot 10^{14}\ W/cm^2$ with different phase- and group velocities is shown. The electrical field of the driving laser at different positions during propagation (colored lines in a)) in the nonlinear medium is shown in b), c) and d). The propagation is in the time frame of the phase velocity of the driving laser. The red highlighted laser cycle in b), c) and d) depicts the region of interest for the HHG-process. The electron trajectories for a return energy of 90 eV for the different electrical driving fields are shown in e). A zoom in (red dashed square) at the time of return is shown in f).*

Ultimately, the different return times are the classically equivalent to a quantum mechanically calculated intrinsic phase $\Phi_{\text{intrinsic}}$, while a coherent superposition is only possible for a phase difference of $\Delta\Phi_{\text{intrinsic}} = \pi$. To achieve this, the electric field of the driving laser must be only minimally deformed during propagation, assuring similar electron trajectories and hence a similar intrinsic phase at each position. In certain cases, this is only possible over a short propagation distance, as is the case in this particular example. As a result, the coherence length (corresponding to the walk-off length) and hence the effective medium length can be significantly shorter than the actual medium length.

This description is an oversimplification of the underlying physics. However, it vividly illustrates the basic concept of the CEP walk-off limit. More precisely, the ionization potential of the target atoms must be considered and the laser does not propagate through a uniform ionized medium, as the ionization is induced by the laser itself. Furthermore, the intrinsic phase during propagation must be calculated, and the return times are only illustrative. However, even if all other parameters (atomic and plasma dispersion as well as the Gouy phase) are optimized for perfect phase matching ($\Delta k = 0$), a coherent build-up is only possible for electron trajectories that have an intrinsic phase difference of $\Delta\Phi_{\text{intrinsic}} \leq \pi$, thus over a finite propagation distance and a reduced coherence length.

## Quantitative Analysis

In the following, a quantitative analysis of the reduced coherence length, that can be interpreted as the walk-off length due to the CEP walk-off is presented and compared to the absorption length. With this knowledge, it will be possible to determine whether experiments can be conducted in the absorption-limited regime. Similar to the argumentation in[24], the walk-off length can be defined as:

$$L_{\text{walk-off}}^{\Delta\Phi=\pi} = \frac{t_{\text{walk-off}}^{\Delta\Phi_{\text{intrinsic}}=\pi}}{|v_{\text{group}}^{-1} - v_{\text{phase}}^{-1}|}, \qquad 6$$

with $v_{\text{group}}$ and $v_{\text{phase}}$ as the group- and phase velocities of the driving laser, respectively. The envelope walk-off time $t_{\text{walk-off}}^{\Delta\Phi_{\text{intrinsic}}=\pi}$ to imprint a phase-shift of $\Delta\Phi_{\text{intrinsic}} = \pi$ on the intrinsic phase is defined as:

$$t_{\text{walk-off}}^{\Delta\Phi_{\text{intrinsic}}=\pi} = \frac{\text{CEP}_{\text{walk-off}}^{\Delta\Phi_{\text{intrinsic}}=\pi}}{\omega_0} \qquad 7$$

with $\text{CEP}_{\text{walk-off}}^{\Delta\Phi_{\text{intrinsic}}=\pi}$ as the tolerable CEP difference to imprint a phase-shift of $\Delta\Phi_{\text{intrinsic}} = \pi$ on the intrinsic phase using a fundamental laser with angular frequency $\omega_0$. The $\text{CEP}_{\text{walk-off}}^{\Delta\Phi_{\text{intrinsic}}=\pi}$ can be obtained using numerical simulations, calculating the intrinsic phase of electrons with equal return energies for different CEP values of the driving laser.

Most importantly, it is crucial to determine whether a process is absorption-limited or CEP walk-off limited. Therefore, a dimensionless value to compare the walk-off length and the absorption length ($L_{\text{abs}}$) can be introduced: $\xi = L_{\text{walk-off}}^{\Delta\Phi_{\text{intrinsic}}=\pi} / L_{\text{abs}}$. If $\xi > 5$ (i.e., $L_{\text{walk-off}}^{\Delta\Phi_{\text{intrinsic}}=\pi} > 5\, L_{\text{abs}}$) the HHG process is absorption-limited[26], and the CEP walk-off can be neglected. However, if $\xi < 5$ (i.e., $L_{\text{walk-off}}^{\Delta\Phi_{\text{intrinsic}}=\pi} < 5\, L_{\text{abs}}$) the absorption limit fundamentally cannot be reached, and the process is CEP walk-off limited.

If the value of $\text{CEP}_{\text{walk-off}}^{\Delta\Phi_{\text{intrinsic}}=\pi}$ is known and a constant ionization level is assumed, $\xi$ can be derived analytically by using the dispersion of the plasma and the neutral atoms. In principle, the Gouy phase also introduces a CEP walk-off of the driving laser. Since the HHG buildup length is limited to roughly one Rayleigh length, the maximum Gouy phase-induced CEP walk-off is $< 0.3\pi$, which is negligible compared to the contributions of the dispersion of the partly ionized medium. The resulting formula for $\xi$ is:

$$\xi = \sigma \cdot \text{CEP}_{\text{walk-off}}^{\Delta\Phi_{\text{intrinsic}}=\pi} \left[ \eta \frac{1}{\omega_0} \frac{e^2}{c_0 \varepsilon_0 m_e} + (1-\eta)\omega_0^2 \frac{1}{c_0 N_{\text{atm}}} \left. \frac{\partial \delta_0(\omega)}{\partial \omega} \right|_{\omega=\omega_0} \right]^{-1}, \qquad 8$$

with the photo ionization cross-section of the generated XUV radiation $\sigma$, the ionization fraction $\eta$, the vacuum speed of light $c_0$, the vacuum permittivity $\varepsilon_0$, the electron mass $m_e$, the number density at standard pressure $N_{\text{atm}}$ and $\delta_0$ being related to the index of refraction at standard conditions as $n_0 = 1 + \delta_0(\omega)$. The first summand in equation 8 is due to the plasma and the second summand relates to the neutral atoms. Note, that $\xi$ is independent on the particle density of the generating medium. Therefore, it can provide general insights into limiting effects using different driving wavelengths, pulse durations and gas types.

Selected Results

This section will discuss selected results of the CEP walk-off simulation, illustrating the dependency of the CEP walk-off on the driving pulse duration and photon energy for selected gases and fundamental wavelengths. The results are shown in Figure 6. Every plot is divided in three parts: (1) absorption limited regime (dark red), where $\xi > 5$, (2) CEP- walk-off limited regime, where $\xi < 5$, and (3) the non-phase matching regime due to the phase matching cutoff (white).

The maps in Figure 6 are calculated using equation 8, while the ionization fraction is set to the critical ionization to incorporate phase matching. However, for the discussion of underlying physics, equation 8 can be further simplified to

$$\xi \propto \frac{\sigma}{\lambda_0 \eta_{\text{crit}}} \cdot \text{CEP}_{\text{walk-off}}^{\Delta\Phi_{\text{intrinsic}}=\pi}. \qquad 9$$

This is due to the fact, that noble gases, commonly used for HHG sources, exhibit minimal dispersion in the visible to infrared spectral regions compared to the plasma dispersion. Consequently, the primary influences on $\xi$ are the critical ionization fraction $\eta_{\text{crit}}$, the absorption cross-section of different gases and photon energies $\sigma$ and the fundamental wavelength $\lambda_0$. Note that the discussion applies to all generating geometries, i.e., waveguide, tight focusing and loose focusing. This is due to the similarities in the wave-vector mismatch caused by the Gouy phase and waveguide dispersion.

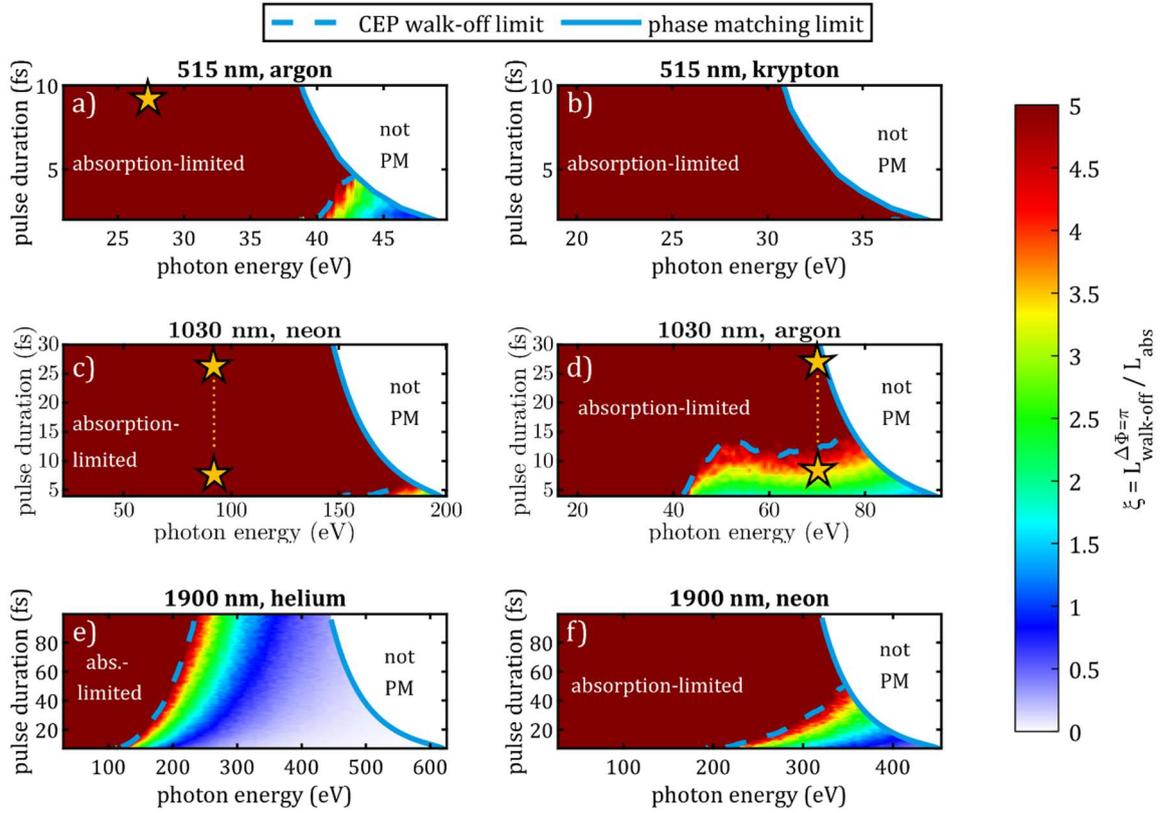

*Figure 6: Calculation of the ratio of the walk-off and absorption length for different fixed gases and fundamental wavelengths and varying XUV photon energies and driving pulse durations. The dashed blue line indicates the border between the CEP- walk-off and absorption limited regimes. The solid blue lines show the phase matching cutoff. The stars mark experiments, discussed in previous sections. Figure adapted from [48].*

In general, regardless of the type of noble gas, the absorption cross-section is larger for lower photon energies[1]. Consequently, absorption-limited HHG is always feasible for photon energies below 40 eV, irrespective of the pulse duration and for driving wavelengths ranging from 500 nm to 2 µm. Additionally, the CEP walk-off is significant only for ultrashort driving pulse durations due to the substantial variation in amplitude between adjacent half-cycles. For longer pulse durations, the shape of the electrical field remains nearly constant over many laser cycles, resulting in a large tolerable $\text{CEP}_{\text{walk-off}}^{\Delta\Phi_{\text{intrinsic}}=\pi}$.

The CEP walk-off is more pronounced for longer driving wavelengths. Specifically, at a wavelength of 1900 nm and using helium, it is fundamentally impossible to reach the absorption-limited regime in the water window. This is due to two main reasons: (1) The same pulse duration corresponds to a different number of cycles for different wavelengths, making long-wavelength drivers more susceptible to CEP walk-off effects at equal pulse durations. (2) Long-wavelength drivers are typically used to generate high photon energies. However, the absorption cross-section decreases with increasing photon energies[1]. This necessitates a longer medium length for absorption-limited HHG, which cannot be achieved due to the CEP walk-off.

It is important to mention, that the results presented only compare whether the HHG process can be conducted in the absorption limit or the CEP walk-off limit. However, this does not imply that HHG in the CEP walk-off limit is always less efficient, as shown in Figure 4.

In Figure 6, the stars represent the various experiments discussed in the previous sections. The figure demonstrates that experiments conducted with a fundamental wavelength of 515 nm, optimized for 21 eV and 26 eV in argon, can always be performed in the absorption-limited regime. Additionally, the fundamental limitation for experiments in neon, using a fundamental wavelength of 1030 nm and targeting a photon energy of 92 eV, is the absorption limit. Consequently, the efficiency scaling law in these experiments follows the single-atom efficiency scaling (equation 3). Conversely, HHG in argon using a fundamental wavelength of 1030 nm and pulse durations shorter than 15 fs is CEP walk-off limited for photon energies ranging from 40 eV up to the phase-matching cutoff of <100 eV. In this scenario, the anticipated increase in conversion efficiency with shorter pulses will be impeded by the CEP walk-off effect.

## 4 Summary and outlook for Watt-class high-harmonic sources

In summary, a theory for the efficiency scaling laws of HHG-based XUV to soft X-ray sources is presented. This includes a scaling of the single-atom dipole moment with the driving pulse duration under phase-matched conditions, demonstrating a quantitative scaling of $\tau^{-\kappa_{\text{single atom}}}$ with $\kappa_{\text{single atom}}$ ranging from 0.6 to 1.1. This scaling law has been validated at a driving wavelength of 515 nm by comparing experiments targeting a photon energy range of <30 eV with different driving pulse durations.

Additionally, a new limitation has been introduced – the CEP walk-off limit. This limitation can significantly reduce the coherence length, thereby hindering the generation of absorption-limited HHG. This issue is particularly severe at high photon energies. Specifically, in the soft X-ray region at photon energies above 300 eV, it is fundamentally impossible to drive the HHG process in the absorption-limited regime. Additional to the driving wavelength-scaling of the single atom response, stating that longer driver wavelengths result in a significantly lower HHG-efficiency[21], this contributes to the very low conversion efficiencies reported for HHG in this spectral region. Ultimately it is nearly impossible to design highly efficient HHG-based X-ray sources with photon energies larger than 1 keV, using a conventional HHG setup.

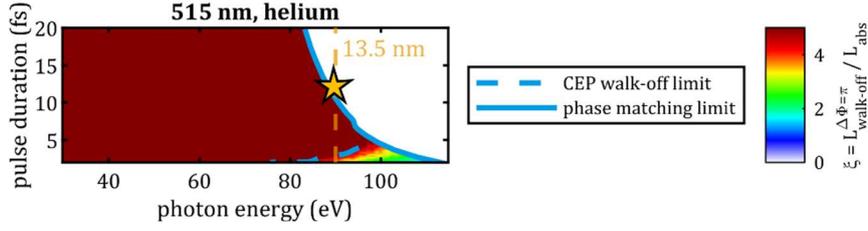

*Figure 7: Ratio of the CEP- walk-off length and the absorption length for 515 nm HHG in helium. The blue lines correspond to the CEP walk-off and the phase matching limit. The orange dashed line highlights a wavelength 13.5 nm. The yellow star indicates parameters with a great potential for highly efficient HHG beamlines at 13.5 nm.*

In contrast, recent advancements in high peak power and high average power laser systems in the visible range open new opportunities for highly efficient HHG beamlines in the XUV at photon energies lower than 100 eV. The wavelength of 515 nm is particularly interesting as it can be produced as the second harmonic of the matured Yb-fiber laser technology. These systems regularly deliver pulses several hundred femtoseconds long with pulse energies of several millijoules and average powers exceeding 1 kW[50]. By generating the second harmonic and using multipass-cell[51] or fiber-based[15] compression, future systems have the potential to achieve more than 500 W average power and pulse durations of less than 20 fs. These parameters are promising for the highly efficient generation of HHG-based 13.5 nm radiation. As shown in Figure 7, 515 nm-driven HHG in helium is always absorption-limited for pulse durations longer than 6 fs, and the phase matching cutoff of 92 eV can be reached with pulses shorter than 25 fs. The combination with the $\lambda^{-6}$ scaling of the single-atom response[21] will enhance efficiency, resulting in theoretical efficiencies of better than $\eta = 10^{-7}$[^1], one order of magnitude higher than state of the art efficiencies for 13.5 nm HHG-based systems[20]. Combined with the potential 500 W average power, this could lead to an average power exceeding 100 µW at 13.5 nm, which is three orders of magnitude higher than current state-of-the-art sources[20,35].

Ultimately, at low photon energies below 30 eV, a conversion efficiency into a single harmonic line as high as $5 \cdot 10^{-4}$ might be achievable[15]. This can be realized through further efforts to produce pulses shorter than 10 fs at a fundamental wavelength of 515 nm with a high average power of several hundred watts—twice as short and ten times higher in average power compared to the current state of the art[15]. This has the potential to boost the efficiency by a factor of 20 (as per equation 4) and could potentially lead to several hundred milliwatts per harmonic line and watt-level average power if multiple harmonic lines are considered.

Finally, the introduced concept paves the way for the development of a new generation of highly efficient XUV and soft X-ray sources. Additionally, more compact, few-watt driving lasers can be utilized, enabling a wider distribution of HHG sources, potentially extending beyond specialized laser laboratories. This advancement could address grand challenges in fields such as health and energy, significantly enhancing existing photon-hungry applications and enabling novel technologies in both fundamental and applied sciences.

[^1]: This has been estimated by utilizing the efficiency of $3 \cdot 10^{-8}$, achieved with an 800 nm, 45 fs and 2 mJ laser for HHG in Helium in a waveguide geometry[20], and applying corrections due to the wavelength scaling of the single atom response (x14), the higher applicable intensity due modal averaging effects of the ionization (x0.25), and the usage of a shorter pulse duration (x2).

# References

1. Henke, B. L., Gullikson, E. M. & Davis, J. C. X-Ray Interactions: Photoabsorption, Scattering, Transmission, and Reflection at E = 50-30,000 eV, Z = 1-92. *Atomic Data and Nuclear Data Tables* **54,** 181–342; 10.1006/adnd.1993.1013 (1993).

2. Gaumnitz, T. *et al.* Streaking of 43-attosecond soft-X-ray pulses generated by a passively CEP-stable mid-infrared driver. *Optics express* **25,** 27506–27518; 10.1364/OE.25.027506 (2017).

3. Mathias, S. *et al.* Ultrafast element-specific magnetization dynamics of complex magnetic materials on a table-top. *Journal of Electron Spectroscopy and Related Phenomena* **189,** 164–170; 10.1016/j.elspec.2012.11.013 (2013).

4. Sakdinawat, A. & Attwood, D. Nanoscale X-ray imaging. *Nature Photon* **4,** 840–848; 10.1038/nphoton.2010.267 (2010).

5. Krausz, F. & Ivanov, M. Attosecond physics. *Rev. Mod. Phys.* **81,** 163–234; 10.1103/RevModPhys.81.163 (2009).

6. Ciappina, M. F. *et al.* Attosecond physics at the nanoscale. *Reports on progress in physics. Physical Society (Great Britain)* **80,** 54401; 10.1088/1361-6633/aa574e (2017).

7. Loetgering, L., Witte, S. & Rothhardt, J. Advances in laboratory-scale ptychography using high harmonic sources Invited. *Optics express* **30,** 4133–4164; 10.1364/OE.443622 (2022).

8. Schoenlein, R. *et al.* Recent advances in ultrafast X-ray sources. *Philosophical transactions. Series A, Mathematical, physical, and engineering sciences* **377,** 20180384; 10.1098/rsta.2018.0384 (2019).

9. Sedigh Rahimabadi, P., Khodaei, M. & Koswattage, K. R. Review on applications of synchrotron-based X-ray techniques in materials characterization. *X-Ray Spectrometry* **49,** 348–373; 10.1002/xrs.3141 (2020).

10. Seddon, E. A. *et al.* Short-wavelength free-electron laser sources and science: a review. *Reports on progress in physics. Physical Society (Great Britain)* **80,** 115901; 10.1088/1361-6633/aa7cca (2017).

11. Chang, Z. *Fundamentals of Attosecond Optics.* 1st ed. (CRC press; Safari, Erscheinungsort nicht ermittelbar, Boston, MA, 2016).

12. Rocca, J. J. Table-top soft x-ray lasers. *Review of Scientific Instruments* **70,** 3799–3827; 10.1063/1.1150041 (1999).

13. Hädrich, S. *et al.* Single-pass high harmonic generation at high repetition rate and photon flux. *J. Phys. B: At. Mol. Opt. Phys.* **49,** 172002; 10.1088/0953-4075/49/17/172002 (2016).

14. Pupeza, I. *et al.* Compact high-repetition-rate source of coherent 100 eV radiation. *Nature Photon* **7,** 608–612; 10.1038/nphoton.2013.156 (2013).

15. Klas, R. *et al.* Ultra-short-pulse high-average-power megahertz-repetition-rate coherent extreme-ultraviolet light source. *PhotoniX* **2**; 10.1186/s43074-021-00028-y (2021).

16. Porat, G. *et al.* Phase-matched extreme-ultraviolet frequency-comb generation. *Nature Photon* **12,** 387–391; 10.1038/s41566-018-0199-z (2018).

17. Grebing, C., Müller, M., Buldt, J., Stark, H. & Limpert, J. Kilowatt-average-power compression of millijoule pulses in a gas-filled multi-pass cell. *Optics letters* **45,** 6250–6253; 10.1364/OL.408998 (2020).

18. Kramer, P. L. *et al.* Enabling high repetition rate nonlinear THz science with a kilowatt-class sub-100 fs laser source. *Opt. Express, OE* **28,** 16951–16967; 10.1364/OE.389653 (2020).

19. Pfaff, Y. *et al.* Nonlinear pulse compression of a 200 mJ and 1 kW ultrafast thin-disk amplifier. *Opt. Express, OE* **31,** 22740–22756; 10.1364/OE.494359 (2023).

20. Ding, C. *et al.* High flux coherent super-continuum soft X-ray source driven by a single-stage, 10mJ, Ti:sapphire amplifier-pumped OPA. *Optics express* **22,** 6194–6202; 10.1364/OE.22.006194 (2014).

21. Shiner, A. D. *et al.* Wavelength scaling of high harmonic generation efficiency. *Physical review letters* **103,** 73902; 10.1103/PhysRevLett.103.073902 (2009).

22. Christov, I. P. *et al.* Nonadiabatic Effects in High-Harmonic Generation with Ultrashort Pulses. *Phys. Rev. Lett.* **77,** 1743–1746; 10.1103/PhysRevLett.77.1743 (1996).

23. Westerberg, S. *et al.* Influence of the laser pulse duration in high-order harmonic generation. *APL Photonics* **10**; 10.1063/5.0272968 (2025).

24. Hernández-García, C. *et al.* Group velocity matching in high-order harmonic generation driven by mid-infrared lasers. *New J. Phys.* **18,** 73031; 10.1088/1367-2630/18/7/073031 (2016).

25. Kroon, D. *et al.* Attosecond pulse walk-off in high-order harmonic generation. *Optics letters* **39,** 2218–2221; 10.1364/OL.39.002218 (2014).

26. Constant, E. *et al.* Optimizing High Harmonic Generation in Absorbing Gases: Model and Experiment. *Phys. Rev. Lett.* **82,** 1668–1671; 10.1103/PhysRevLett.82.1668 (1999).

27. Lewenstein, M., Balcou, P., Ivanov, M. Y., L'Huillier, A. & Corkum, P. B. Theory of high-harmonic generation by low-frequency laser fields. *Phys. Rev. A* **49,** 2117–2132; 10.1103/PhysRevA.49.2117 (1994).

28. Ishikawa, K. L., Schiessl, K., Persson, E. & Burgdörfer, J. Fine-scale oscillations in the wavelength and intensity dependence of high-order harmonic generation: Connection with channel closings. *Phys. Rev. A* **79**; 10.1103/PhysRevA.79.033411 (2009).

29. Zaïr, A. *et al.* Quantum path interferences in high-order harmonic generation. *Physical review letters* **100,** 143902; 10.1103/PhysRevLett.100.143902 (2008).

30. Kazamias, S. *et al.* Pressure-induced phase matching in high-order harmonic generation. *Phys. Rev. A* **83**; 10.1103/PhysRevA.83.063405 (2011).

31. Weissenbilder, R. *et al.* How to optimize high-order harmonic generation in gases. *Nat Rev Phys* **4,** 713–722; 10.1038/s42254-022-00522-7 (2022).

32. Popmintchev, T., Chen, M.-C., Arpin, P., Murnane, M. M. & Kapteyn, H. C. The attosecond nonlinear optics of bright coherent X-ray generation. *Nature Photon* **4,** 822–832; 10.1038/nphoton.2010.256 (2010).

33. Minneker, B., Klas, R., Rothhardt, J. & Fritzsche, S. Critical Laser Intensity of Phase-Matched High-Order Harmonic Generation in Noble Gases. *Photonics* **10,** 24; 10.3390/photonics10010024 (2023).

34. Corkum, P. B. Plasma perspective on strong field multiphoton ionization. *Phys. Rev. Lett.* **71,** 1994–1997; 10.1103/PhysRevLett.71.1994 (1993).

35. Klas, R., Eschen, W., Kirsche, A., Rothhardt, J. & Limpert, J. Generation of coherent broadband high photon flux continua in the XUV with a sub-two-cycle fiber laser. *Optics express* **28,** 6188–6196; 10.1364/OE.28.006188 (2020).

36. Ammosov, M. V., Delone, N. B. & Krainov, V. P. Tunnel Ionization Of Complex Atoms And Atomic Ions In Electromagnetic Field. *Sov. Phys. JETP* **64,** 138; 10.1117/12.938695 (1986).

37. Perelomov, A. M. and Popov, V. S. and Terent'ev, M. V. Ionization of atoms in an alternating electrical field. *Sov. Phys. JETP* **5,** 924–934 (1966).

38. Zhou, J., Peatross, J., Murnane, M. M., Kapteyn, H. C. & Christov, I. P. Enhanced high-harmonic generation using 25 fs laser pulses. *Phys. Rev. Lett.* **76,** 752–755; 10.1103/PhysRevLett.76.752 (1996).

39. Klas, R. *et al.* Table-top milliwatt-class extreme ultraviolet high harmonic light source. *Optica* **3,** 1167; 10.1364/OPTICA.3.001167 (2016).

40. Klas, R., Kirsche, A., Tschernajew, M., Rothhardt, J. & Limpert, J. Annular beam driven high harmonic generation for high flux coherent XUV and soft X-ray radiation. *Optics express* **26,** 19318–19327; 10.1364/OE.26.019318 (2018).

41. Hilbert, V., Tschernajew, M., Klas, R., Limpert, J. & Rothhardt, J. A compact, turnkey, narrow-bandwidth, tunable, and high-photon-flux extreme ultraviolet source. *AIP Advances* **10**; 10.1063/1.5133154 (2020).

42. Comby, A. *et al.* Cascaded harmonic generation from a fiber laser: a milliwatt XUV source. *Optics express* **27,** 20383–20396; 10.1364/OE.27.020383 (2019).

43. Li, K. *et al.* Tabletop ptychographic imaging system with a 515 nm laser driven high-order harmonic source. *Optics and Lasers in Engineering* **176,** 108105; 10.1016/j.optlaseng.2024.108105 (2024).

44. Skruszewicz, S. *et al.* Table-top interferometry on extreme time and wavelength scales. *Optics express* **29,** 40333–40344; 10.1364/OE.446563 (2021).

45. Hädrich, S. *et al.* Scalability of components for kW-level average power few-cycle lasers. *Applied optics* **55,** 1636–1640; 10.1364/AO.55.001636 (2016).

46. Heyl, C. M., Güdde, J., L'Huillier, A. & Höfer, U. High-order harmonic generation with µJ laser pulses at high repetition rates. *J. Phys. B: At. Mol. Opt. Phys.* **45,** 74020; 10.1088/0953-4075/45/7/074020 (2012).

47. Rothhardt, J. *et al.* Absorption-limited and phase-matched high harmonic generation in the tight focusing regime. *New J. Phys.* **16,** 33022; 10.1088/1367-2630/16/3/033022 (2014).

48. Klas, R. Efficiency Scaling of High Harmonic Generation using Ultrashort Fiber Lasers. Dissertation. Friedrich-Schiller-University Jena, 2021.

49. Gebhardt, M. *et al.* Bright, high-repetition-rate water window soft X-ray source enabled by non-linear pulse self-compression in an antiresonant hollow-core fibre. *Light, science & applications* **10,** 36; 10.1038/s41377-021-00477-x (2021).

50. Stark, H., Buldt, J., Müller, M., Klenke, A. & Limpert, J. 1 kW, 10 mJ, 120 fs coherently combined fiber CPA laser system. *Optics letters* **46,** 969–972; 10.1364/OL.417032 (2021).

51. Karst, M. *et al.* 22-W average power high pulse energy multipass-cell-based post-compression in the green spectral range. *Optics letters* **48,** 1300–1303; 10.1364/OL.482600 (2023).